\documentstyle[12pt]{article}
\begin{document}
\title{Imprints of Discrete Space Time - A Brief Note}
\author{B.G. Sidharth\\Centre for Applicable Mathematics \& Computer Sciences\\
B.M. Birla Science Centre, Hyderabad 500 063}
\date{}
\maketitle
\begin{abstract}
We point out that the observed decay mode of the pion and the Kaon decay puzzle
are really imprints of discrete micro space-time.
\end{abstract}
In recent years ideas of discrete space time have been revived through the
work of several scholars and by the author within the context of Kerr-Newman Black
Hole type formulation of the electron\cite{r1}-\cite{r5}. Further, even more recently
this has been considered in the
context of a stochastic underpinning\cite{r6,r7}. Let us now consider two of the imprints
that such discrete space time would have.\\
First we consider the case of the
neutral pion. Within the framework of the Kerr-Newman metric type formulation
referred to above, it is possible to recover the usual picture of a pion as a
quark-anti quark bound state \cite{r8,r9}, though equally well we could think
of it as an electron-positron bound state also\cite{r4,r10}. In this case we
have,
\begin{equation}
\frac{mv^2}{r} = \frac{e^2}{r^2}\label{e1}
\end{equation}
Consistently with the above formulation, if we take $v = c$ from (\ref{e1}) we get the correct Compton wavelength
$l_\pi = r$ of the pion.\\
However this appears to go against the fact that there would be pair annihilation
with the release of two photons. However if we consider discrete space time,
the situation would be different. In this case the Schrodinger equation
\begin{equation}
H \psi = E \psi\label{e2}
\end{equation}
where $H$ contains the above Coulumb interaction could be written, in terms
of the space and time separated wave function components as (Cf. also ref.\cite{r2}),
\begin{equation}
H\psi = E \phi T = \phi \imath \hbar [\frac{T(t-\tau)-T}{\tau}]\label{e3}
\end{equation}
where $\tau$ is the minimum time cut off which in the above work has been taken to be the Compton
time (Cf.refs.\cite{r4} and \cite{r5}). If, as usual we let $T = exp (irt)$ we get
\begin{equation}
E = -\frac{2\hbar}{\tau} sin \frac{\tau r}{2}\label{e4}
\end{equation}
(\ref{e4}) shows that if,
\begin{equation}
| E | < \frac{2\hbar}{\tau}\label{e5}
\end{equation}
holds then there are stable bound states. Indeed inequality (\ref{e5}) holds
good when $\tau$ is the Compton time and $E$ is the total energy $mc^2$. Even if
inequality (\ref{e5}) is reversed, there are decaying states which are relatively
stable around the cut off energy $\frac{2\hbar}{\tau}$.\\
This is the explanation for treating the pion as a bound state of an electron
and a positron, as indeed is borne out by its decay mode.
The situation is similar to
the case of Bohr orbits-- there also the electrons would according to classical
ideas have collapsed into the nucleus and the atoms would have disappeared. In
this case it is the discrete nature of space time which enables the pion to be
a bound state as described by (\ref{e1}).\\
Another imprint of discrete space time can be found in the Kaon decay puzzle,
as pointed out by the author\cite{r11}. There also  we have equations like
(\ref{e2}) and (\ref{e3}) above, with the energy
term being given by $E(1 + i)$, due to the fact that space time is quantized.
Not only is the fact that the imaginary and real parts of the energy are of
the same order is borne out but as pointed out in\cite{r11} this also
explains the recently observed\cite{r12} decay and violation of the time reversal
symmetry which in the words of Penrose\cite{r13}, "the tiny fact of an
almost completely hidden time-asymmetry seems genuinely to be present in the
$K^0$-decay. It is hard to believe that nature is not, so to speak, trying to
tell something through the results of this delicate and beautiful experiment."\\
From an intuitive point of view, the above should not be surprising because time
reversal symmetry is based on a space time continuum and is no longer obvious
if space time were discrete.

\end{document}